\documentclass[12pt]{iopart}

\begin{document}

\title{Weakened linearity for quantum fields}
\author{Peter Morgan}
\address{Physics Department, Yale University, CT 06520.}
\ead{peter.w.morgan@yale.edu}

\begin{abstract}
There are still no interacting models of the Wightman axioms, suggesting that the
axioms are too tightly drawn.
Here a weakening of linearity for quantum fields is proposed, with the algebra still
linear but with the quantum fields no longer required to be tempered distributions,
allowing explicit interacting quantum field models.
Interacting quantum fields should be understood to be nonlinear quantum fields in this
sense, because a set of effective field theories encodes a dependence on the energy scale
of measurement --- which is a nontrivial property of the test functions --- so that
correlation functions are implicitly nonlinear functions of test functions in the
conventional formalism.
In Local Quantum Physics terms, the algebraic models constructed here do not satisfy
the additivity property.
Finite nonlinear deformations of quantized electromagnetism are constructed as examples.
\end{abstract}

\submitto{\JPA}
\maketitle

\newcommand\Half{{\frac{1}{2}}}
\newcommand\Intd{{\mathrm{d}}}
\newcommand\eqN{{\,\stackrel{\mathrm{N}}{=}\,}}
\newcommand\PP[1]{{(\hspace{-.27em}(#1)\hspace{-.27em})}}
\newcommand\PPs[1]{{(\hspace{-.4em}(#1)\hspace{-.4em})}}
\newcommand\RR {{\mathrm{I\hspace{-.1em}R}}}
\newcommand\CC{{{\rm C}\kern -0.5em 
          \vrule width 0.05em height 0.65em depth -0.03em
          \kern 0.45em}}
\newcommand\kT{{{\mathsf{k_B}} T}}
\newcommand\RE{{\mathrm{Re}}}
\newcommand\IM{{\mathrm{Im}}}

\section{Introduction}
The free Klein-Gordon quantum field is an operator valued \emph{linear} map from a
suitable space of functions, $\hat\phi:f\mapsto\hat\phi_f$.
We will take $f$ to be from a Schwartz space of functions\cite[\S II.1.2]{Haag}, so that
$f(x)$ is infinitely often differentiable and decreases as well as its derivatives faster
than any power as $x$ moves to infinity in any direction.
For the free Klein-Gordon quantum field, $\hat\phi$ is then a tempered distribution.
This is the linearity we will weaken: we will allow the operator valued map
$\hat\phi:f\mapsto\hat\phi_f$ to be nonlinear, so that the linear operators $\hat\phi_f$,
$\hat\phi_g$ and $\hat\phi_{f+g}$ will in general not satisfy the linear dependence
$\hat\phi_f+\hat\phi_g = \hat\phi_{f+g}$.
With this weakening, we cannot take a quantum field to be an operator-valued
distribution $\hat\phi(x)$, we will be concerned \emph{only} with operators $\hat\phi_f$.
Note, however, that allowing $\hat\phi$ to be nonlinear does not weaken the linearity
of the algebra generated by the operators $\hat\phi_f$, and we will be able to construct
a linear Hilbert space representation of the algebra of observables.
The construction here is thus different from the nonlinear relativistic approach of Kibble,
for example, who introduces a nonlinear Hamiltonian operator\cite{Kibble}.

The nonlinear quantum fields constructed here do not satisfy the additivity property of
Local Quantum Physics \cite[\textit{Axiom} \textbf{B}, \S III.1]{Haag}.
This axiom requires that two algebras of observables, associated with regions $\mathcal{O}_1$
and $\mathcal{O}_2$ in space-time, together generate the algebra of observables associated
with their union, $\mathcal{A}(\mathcal{O}_1\cup\mathcal{O}_2)=
       \mathcal{A}(\mathcal{O}_1)\vee\mathcal{A}(\mathcal{O}_2)$, but this is generally not
possible if, for $f$ and $g$ with support in $\mathcal{O}_1$ and $\mathcal{O}_2$ respectively,
$\hat\phi_{f+g}\not=\hat\phi_f+\hat\phi_g$.
The construction of this paper therefore casts some doubt on the necessity of the
additivity property as an axiom of quantum field theory.

Locality and Lorentz covariance, however, will be preserved absolutely.
The algebraic structure of a free linear quantum field is given by the hermitian inner product
corresponding to the commutator, $[\hat a^{ }_g, \hat a_f^\dagger]=(f,g)$, with
$\hat\phi_f=\hat a^{ }_f+\hat a_f^\dagger$.
A free linear quantum field is local just because $[\hat\phi_f,\hat\phi_g]=(g,f)-(f,g)$ is zero
whenever the test functions $f$ and $g$ have space-like separated supports.
Nonlinearity will be introduced in two ways, firstly by the simple expedient of taking the
commutator to be a sum of a number of inner products such as, for example, without worrying here
about constants,
\begin{eqnarray}
  [\hat a^{ }_g, \hat a_f^\dagger]&=&\xi(f,g)=(f,g)+(f+f^2,g+g^2)+(f^2,g^2)+\left(f(f,f),g(g,g)\right)+
\cr    &&\qquad 
                        (f+\partial_\mu f\partial^\mu f,g+\partial_\mu g\partial^\mu g)+...,
\end{eqnarray}
which will result in a local nonlinear quantum field just because invariant polynomials in the
field and its derivatives such as $f^n(x)=[f(x)]^n$ or $\partial_\mu f\partial^\mu f$ have support
contained in $\mathrm{Supp}(f)$.
$[\hat\phi_f,\hat\phi_g]=\xi(g,f)-\xi(f,g)$ is zero, as for the free field, whenever the test
functions $f$ and $g$ have space-like separated supports.
The constraints of locality, positive semi-definiteness and Lorentz invariance on the form of
$\xi(f,g)$ are satisfied by many models, and it will turn out to be as easy to construct a
vacuum state over this algebra as over the linear free field, allowing the GNS construction
of a Hilbert space.

Secondly, we can deform the simple relationship $\hat\phi_f=\hat a^{ }_f+\hat a_f^\dagger$, setting
$\hat\phi_f$ to be an arbitrary self-adjoint operator-valued function of $\hat a^{ }_f+\hat a_f^\dagger$,
$\hat a^{ }_{\mathcal{P}_i[f]}+\hat a_{\mathcal{P}_i[f]}^\dagger$, ...,
\begin{equation}
    \hat\phi_f=\hat F(\hat a^{ }_f+\hat a_f^\dagger,\hat a^{ }_{\mathcal{P}_1[f]}+\hat a_{\mathcal{P}_1[f]}^\dagger,
      \hat a^{ }_{\mathcal{P}_2[f]}+\hat a_{\mathcal{P}_2[f]}^\dagger, X_1(f),X_2(f), ...),
\end{equation}
where $\mathrm{Supp}(\mathcal{P}_i[f])\subseteq\mathrm{Supp}(f)$, and $X_i(f)$ are arbitrary
Poincar\'e invariant scalar functions of $f$ --- microcausality is satisfied whatever such
scalar functions are introduced.
In the general case this deformation is quite nontrivial, more general than a nonlinear
coordinate transformation.

The energy scale of an experiment is essentially a pragmatic matter that is obvious to
an experimenter: an experiment deals with phonons on a lattice, with atomic energy levels,
with nuclear energy levels, etc., without an exact explicit discussion being necessary,
and we can choose the cutoff appropriately for a given experiment without too much
detailed concern.
From a quantum field perspective, however, the energy scale of an experiment is a very
non-detailed measure of the structure of the test functions involved in its description: if a test
function appropriate to a description of an experiment determines an effective real-space length
scale, or if the fourier transform of the same (or another) test function is concentrated at
a particular energy scale, then such scales pragmatically determine what effective field model
we use.
Hence, there is a \emph{prima facie} case that the correlation functions of a quantum field
are nonlinearly determined by properties of the test functions that describe an experiment,
because the test functions are involved in an explicit description of correlation functions
\emph{not only} by smearing, so that interacting quantum fields should be understood to be
nonlinear quantum fields.
This significantly reconceptualizes our understanding of interacting quantum fields.

We will not here concern ourselves with the Hamiltonian operators of the theories we discuss,
because the Hamiltonian is a global (non)observable, so that any constraint on it is essentially
theoretical.
Additionally, the Hamiltonian is inessential to the algebraic constructions of quantum field
theories given here.
Instead, we will take $n$-measurement correlation functions to be the observables of the
theory, with empirical adequacy achieved if a theory can accurately model experimental
correlations.

Section \ref{FreeFieldPreliminaries} first discusses free quantum fields, then section
\ref{WeakenedLinearity1} introduces a large class of models that weaken the linearity of the
quantum field by the introduction of a nonlinear inner product, and section
\ref{WeakenedLinearity2} discusses the introduction of a nonlinear map between $\hat\phi_f$
and creation and annihilation operators.
Section \ref{EMfield} applies the methods of section \ref{WeakenedLinearity1} to an
electromagnetic field, leaving the application of the methods of section \ref{WeakenedLinearity2}
to the future.

\section{Free field preliminaries}\label{FreeFieldPreliminaries}
A simple way to construct the free Klein-Gordon quantum field \cite{MorganCKG} is to project
$\hat\phi_f$ into two parts, $\hat\phi_f=\hat a^{ }_f+\hat a^\dagger_f$, and specify
the algebraic properties of $\hat a^\dagger_f$ and $\hat a^{ }_f$ by the commutation relations
\begin{equation}\label{FreeCommutationRelations}
  \Bigl[\hat a^{ }_g,\hat a^\dagger_f\Bigr]=(f,g),\qquad \Bigl[\hat a^{ }_g,\hat a^{ }_f\Bigr]=0.
\end{equation}
The manifestly Poincar\'e invariant hermitian inner product $(f,g)$ is given by
\begin{equation}\label{ScalarInnerProduct}
  (f,g) = \hbar\int \frac{\Intd^4k}{(2\pi)^4}
                    2\pi\delta(k^\mu k_\mu-m^2)\theta(k_0)\tilde f^*(k)\tilde g(k).
\end{equation}
This fixes the algebraic structure of the observables $\hat\phi_f$,
$[\hat\phi_f,\hat\phi_g] = i\omega(f,g)$, where $\omega(f,g)=i((f,g)-(g,f))=-\omega(g,f)$.
Note that the self-adjoint operators $\hat\phi'_f=i(\hat a^{ }_f-\hat a^\dagger_f)$ are taken
\emph{not} to be observable (if they were observable then we would be able to send messages
faster than light because $[\hat\phi'_f,\hat\phi_g]=i((g,f)+(f,g))$ is non-zero when $f$
and $g$ have space-like separated supports\footnote{We can eliminate the creation and
annihilation operators (which are too prominent in many presentations of the free quantized
Klein-Gordon field), by presenting the algebra directly as $[\hat\phi_f,\hat\phi_g] = i\omega(f,g)$
and presenting the vacuum state using the generating function
$\left<0\right|e^{i\lambda\hat\phi_f}\left|0\right>=
e^{-\Half\lambda^2(f,f)}$, with $\omega(f,g)$ and $(f,g)$ defined as above.
Together these are as sufficient to fix the Wightman functions of the theory as the
construction in the main text is.}).
The vacuum expectation values are fixed by the trivial action of the operators $\hat a^{ }_f$ on
the vacuum state, $\hat a^{ }_f\left|0\right>=0$, and the normalization
$\left<0\right|\!\left.0\right>=1$.
To compute any vacuum expectation value, apply the commutation relations above repeatedly,
eliminating any terms in which $\hat a^{ }_f\left|0\right>$ or $\left<0\right|\hat a^\dagger_f$
appear, until we obtain a number by finally applying $\left<0\right|\!\left.0\right>=1$.
For example, $\left<0\right|\hat\phi_f\hat\phi_g\left|0\right>=
\left<0\right|\hat a^{ }_f\hat a^\dagger_g\left|0\right>=
\left<0\right|((g,f)+\hat a^\dagger_g\hat a^{ }_f)\left|0\right>=(g,f)$.

The commutator algebra and the specification of the vacuum state fix the
Wightman functions of the theory at all times, which effectively encodes all
dynamical information, so that a Hamiltonian and Lagrangian are superfluous in this approach
to quantum fields.
Since the algebra and the definition of the vacuum are the only structures in this
approach, those are what we have to deform to create an interacting field theory.

The free field algebra determines that the probability density associated with
an observable $\hat\phi_f$ in the vacuum state is Gaussian.
The characteristic function can be computed as
$\left<0\right|e^{i\lambda\hat\phi_f}\left|0\right>=
  e^{-\Half\lambda^2(f,f)}$ by applying a Baker-Campbell-Hausdorff formula,
leading to the probability density
$\frac{1}{\sqrt{2\pi(f,f)}}\exp{\left(-\frac{x^2}{2(f,f)}\right)}$, which is well-defined
if we take $f$ to be a Schwartz space function, but not if we take $f$ to be a point-like
delta function.
In a similar way, we can compute the joint quasiprobability density associated with
two observables $\hat\phi_f$ and $\hat\phi_g$ in the vacuum state, which is also Gaussian.
The characteristic function is
$\left<0\right|e^{i\lambda\hat\phi_f+i\mu\hat\phi_g}\left|0\right>=
  e^{-\Half(\lambda f+\mu g,\lambda f+\mu g)}=
  \exp{\left[-\Half\left[\lambda^2(f,f)+2\lambda\mu\textrm{Re}(f,g)+\mu^2(g,g)\right]\right]}$,
leading to the quasiprobability density
\begin{equation}
  \frac{\exp{\left(-\Half\frac{x^2(g,g)-2xy\mathrm{Re}(f,g)+y^2(g,g)}
                              {(f,f)(g,g)-\left|\mathrm{Re}(f,g)\right|^2}\right)}}
       {2\pi\sqrt{(f,f)(g,g)-\left|\mathrm{Re}(f,g)\right|^2}}.
\end{equation}
Note that this quasiprobability is independent of the imaginary parts of $(f,g)$.
Finally for the vacuum state, for a set of observables $\{\hat\phi_{f_j}\}$
we obtain a characteristic function
$\left<0\right|e^{i\sum_j \lambda_j\hat\phi_{f_j}}\left|0\right>=
  e^{-\Half\underline{\lambda}^T F\underline{\lambda}}$, where the matrix $F_{ij}=\mathrm{Re}(f_i,f_j)$
describes the relative geometry of the $n$ joint measurements for the purposes of the free
field theory, leading to the $n$-measurement joint quasiprobability density
\begin{equation}\label{vacuumProb}
  \frac{e^{-\Half \underline{x}^T F^{-1}\underline{x}}}
       {\sqrt{(2\pi)^n \mathrm{det}(F)}}.
\end{equation}
The singular condition $\mathrm{det}(F)=0$ is fairly innocuous, since it is the expectation
values that are significant rather than any characteristic functions that can be used to
generate them.

For the non-vacuum state $\hat a^\dagger_g\left|0\right>/\sqrt{(g,g)}$ and a set of observables
$\{\hat\phi_{f_j}\}$, we obtain a characteristic function
$\left<0\right|\hat a^{ }_g e^{i\sum_j \lambda_j\hat\phi_{f_j}}\hat a^\dagger_g\left|0\right>/(g,g)=
  (1-|\underline{\lambda}.\underline{S}|^2)e^{-\Half\underline{\lambda}^T F\underline{\lambda}}$,
where $S_i=(f_i,g)/\sqrt{(g,g)}$ describes the relation between the state preparation and the
chosen measurements.
This leads to the $n$-measurement joint quasiprobability density
\begin{equation}\label{vacuumplus1Prob}
  \left[|\underline{x}^T F^{-1}\underline{S}|^2+(1-\underline{S}^\dagger F^{-1}\underline{S})\right]
       \frac{e^{-\Half \underline{x}^T F^{-1}\underline{x}}}
            {\sqrt{(2\pi)^n \mathrm{det}(F)}}.
\end{equation}
The imaginary parts of $(f_i,g)$ contribute to equation (\ref{vacuumplus1Prob}), which
consequently may be not positive
semi-definite\footnote{If instead of the inner product of equation (\ref{ScalarInnerProduct}), we use
the real form $(f,g)+(g,f)$, we still obtain a quantum field theory, but it is classical in the sense
that $[\hat\phi_f,\hat\phi_g]=0$ whatever the space-time relationship between $f$ and $g$, and
equation (\ref{vacuumplus1Prob}) is accordingly positive semi-definite.
For a comparable perspective on the relationship between random fields and quantum fields see
\cite{MorganCKG}.}.
It is straightforward, but progressively more time-consuming, to compute $n$-measurement
joint quasiprobability densities for higher states, which introduce increasing deviations
from a Gaussian distribution.
We can in principle also compute probability densities straightforwardly for higher order
observables such as $\hat\phi_{f_1}\hat\phi_{f_2}+\hat\phi_{f_2}\hat\phi_{f_1}$.

The intention of this rather lengthy elementary discussion of characteristic functions and
quasiprobabilities is to give some sense of how we can compute empirically relevant results quite
effectively by only considering the relations between explicit measurement and state descriptions
without ever considering operator-valued distributions $\hat\phi(x)$.
We have exclusively used inner products between the functions $f_i$ and $g$ that were used
above to construct measurements and states.
Using test functions universally has the useful effect of ensuring manifest Poincar\'e invariance
of the resulting formalism very straightforwardly.
Note that we have used the term ``$n$-measurement'' correlations instead of ``$n$-point'',
because we never measure anything at a point, and the idealization of point-like measurements
will become impossible when we introduce nonlinearity.
All calculations involve only Schwartz space functions, which are much easier to manipulate
than distributions, in particular because Schwartz space is closed under multiplication.
In a simple-minded way, it is arguable that the infinities profusely generated by the
conventional perturbation of free quantum fields are caused by the introduction of higher
than quadratic products of distributions.

In more abstract terms, for free fields the properties of the vacuum state define a state
$\varphi_0:A\mapsto \left<0\right|A\left|0\right>$ over the $\star$-algebra $\mathcal{A}$
generated by a finite number of creation and annihilation operators, a linear map satisfying
$\varphi_0(A^\dagger)=\overline{\varphi_0(A)}$, $\varphi_0(A^\dagger A)\ge 0$, $\varphi_0(1)=1$,
which allows the Gelfand-Naimark-Segal construction of a pre-Hilbert space acted on by
$\mathcal{A}$, which can be closed in the norm to obtain a Hilbert space
$\mathcal{H}_{\varphi_0}$ (see Haag\cite[\S III.2]{Haag}).

For free fields, $\varphi_0(A)=\left<0\right|A\left|0\right>$ satisfies
$\varphi_0(A^\dagger A)=\left<0\right|A^\dagger A\left|0\right>\ge 0$ because
\begin{equation}\label{AlgebraIP}
  \left<0\right|\Bigg[\prod_{k=1}^K \hat a^{ }_{f_k}\Bigg]
         \Bigg[\prod_{j=1}^J \hat a^\dagger_{g_j}\Bigg]\left|0\right>=\delta_{J,K}\mathrm{per}[(g_j,f_k)],
\end{equation}
where $\mathrm{per}[(g_j,f_k)]$ is the \textbf{\textsf{permanent}}\footnote{The permanent of a
$K\times K$ matrix $M$ is a sum over the symmetric group,
$\mathrm{per}(M)=\sum_{\sigma\in S_K} M_{1\sigma(1)}M_{2\sigma(2)}...M_{K\sigma(K)}$.
This is the determinant without the sign of the permutation.
The normalized permanent $\mathrm{per}[(g_j,g_k)]/\prod_{i=1}^K(g_i,g_i)$ of a complex
hermitian positive semi-definite matrix that is generated using inner products $(g_j,g_k)$
measures how close the $K$ functions $g_i$ are to being parallel, independently of the
relative lengths $(g_i,g_i)$ of the functions, except in the singular case when
$\prod_{i=1}^K(g_i,g_i)=0$.
If the functions are all parallel, the normalized permanent is $K!$; if they are all orthogonal,
the normalized permanent is $1$.
Comparably, the normalized determinant is zero if \emph{any} subset of the functions is
linearly dependent; if all the functions are orthogonal the normalized determinant is $1$.}
of the $K\times K$ complex matrix $(g_j,f_k)$.
It is well-known\cite{Minc,MarcusNewman} that
\begin{eqnarray}
  &&\mathcal{S}^{\otimes K}\!\times\!\mathcal{S}^{\otimes K}\rightarrow \CC;
    \quad(g_1\otimes ...\otimes g_K,f_1\otimes ...\otimes f_K)\mapsto
      \mathrm{per}[(g_j,f_k)],
\end{eqnarray}
is a complex hermitian positive semi-definite inner product on the symmetrized tensor product
space $\mathcal{S}^{\otimes K}$, so that equation (\ref{AlgebraIP}) defines a complex hermitian
positive semi-definite inner product on a direct sum of symmetrized tensor product spaces.

Any operator constructed as a multinomial in $\hat\phi_{f_i}$ is not in the algebra
$\mathcal{B}(\mathcal{H}_{\varphi_0})$ of bounded observables acting on $\mathcal{H}_{\varphi_0}$,
so we generally have to pay attention to the domain of $A\in\mathcal{A}$.
The insistence on at least a Banach $\star$-algebra structure for the algebra of observables is
useful for analysis (allowing, for example, the extension of the action of the algebra of observables
to the Hilbert space $\mathcal{H}_{\varphi_0}$), but for constructive calculations of expectation
values, characteristic functions, and probability distributions in particular states, as above,
if $\left<\psi\right|A\left|\psi\right>$ is finite for a normalized vector
$\left|\psi\right>\in\mathcal{H}_{\varphi_0}$ then we can interpret $A$ as an observable for
that state.
This is a nontrivial extension of the pre-Hilbert space because, for example, the normalized
vector $e^{\hat a^\dagger_g}\left|0\right>/\sqrt{e^{(g,g)}}$ gives us a finite state
over $\mathcal{A}$.
As well as extending the pre-Hilbert space, we have already implicitly extended the algebra
$\mathcal{A}$ by using $\left<0\right|e^{i\lambda\hat\phi_f}\left|0\right>$ above as a
characteristic function, since $e^{i\lambda\hat\phi_f}$ is not a polynomial in the field.

\section{Weakened linearity I}\label{WeakenedLinearity1}
Suppose now that we replace equation (\ref{FreeCommutationRelations}) by a commutation
relation that depends nonlinearly on $f$ and $g$,
\begin{equation}\label{NonlinearCommutationRelations}
  \Bigl[\hat a^{ }_g,\hat a^\dagger_f\Bigr]=\xi(f,g),\qquad \Bigl[\hat a^{ }_g,\hat a^{ }_f\Bigr]=0,
\end{equation}
where $\xi(f,g)$ must be complex hermitian positive semi-definite on Schwartz space (in the
sense that the matrix $\xi(f_i,f_j)$ is complex hermitian positive semi-definite for any finite set
of Schwartz space functions $\{f_i\}$).
We will call $\xi(f,g)$ a ``nonlinear inner product''; the term ``inner product'' historically
indicates a sesquilinear form, so we will always be explicit about nonlinearity.
The operator valued map $\hat\phi:f\mapsto\hat\phi_f$ cannot be linear if $\xi(f,g)$ is nonlinear.
The algebra $\mathcal{A}_d$ generated by $\hat\phi_f$ is still linear, but the linear dependence
$\hat\phi_f+\hat\phi_g=\hat\phi_{f+g}$ generally does not hold.

Essentially, for any set of vectors $\{g_i\}$ used to construct an operator in the deformed
free field algebra, we obtain a complex hermitian positive semi-definite matrix $\xi(g_i,g_j)$.
As a complex hermitian positive semi-definite matrix, it is a Gram matrix based on some other
functions $\{f_i\}$ chosen so that $(f_i,f_j)=\xi(g_i,g_j)$.
The action of the vacuum state on an operator $A^\dagger A$ in $\mathcal{A}_d$ that is constructed
using $\{\hat a^\dagger_{g_i}\}$ is positive semi-definite, therefore, just because the action of the
vacuum state on an operator constructed in the same way in $\mathcal{A}$ using
$\{\hat a^\dagger_{f_i}\}$ is positive semi-definite.

To ensure locality,
\begin{equation}
  [\hat\phi_f,\hat\phi_g]=\xi(g,f)-\xi(f,g),
\end{equation}
must be zero when $f$ and $g$ have space-like separated supports.
There is a wide range of possibilities for $\xi(f,g)$: we can use the sum of any number
of complex hermitian positive semi-definite inner products such as
\begin{equation}
  (f,g),\,(f+f^2,g+g^2),\,(f^2,g^2),\, ...,\,(f^n,g^n),\,...,
\end{equation}
just because the sum of positive semi-definite matrices is positive semi-definite.
All these terms satisfy locality because $f^n$ has the same support as $f$, so that,
for example, $\omega(f^n,g^n)$ is zero if $f$ and $g$ have space-like separated support.
We can also introduce invariant polynomials in derivatives of the field,
such as $\partial_\mu f\partial^\mu f$, which again have the same support as $f$.
Furthermore, we need not restrict ourselves to one inner product $(f,g)$, we can introduce
different mass Poincar\'e invariant inner products for different invariant polynomials in
the field and its derivatives.
If the free quantum field is a 4-vector or other nontrivial representation space of the
Lorentz group, ``$f^n$'', perhaps contracted in some way, will usually require a
different inner product than $f$ (see section \ref{EMfield} for a concrete example).
In general, $\xi(f,g)$ can be a sum
\begin{equation}
  \xi(f,g)=\sum_i (\mathcal{P}_i[f],\mathcal{P}_i[g])_i
\end{equation}
for a list of local functionals $\mathcal{P}_i$, satisfying
$\mathrm{Supp}(\mathcal{P}_i[f])\subseteq\mathrm{Supp}(f)$, and a list of linear inner products
$(\cdot,\cdot)_i$.

That we cannot in general expect the linear dependencies
$\hat\phi_f+\hat\phi_g=\hat\phi_{f+g}$ and $\hat\phi_{\lambda f}\not=\lambda\hat\phi_f$
to hold requires a fresh understanding of what we do when we describe a measurement using a
function $f+g$ or $\lambda f$, which we must derive from the mathematical structure of the
nonlinear inner product.
In the linear case, we can imagine in folk terms that when we use the operator $\hat\phi_f$ we
are asking how much $f$ ``resonates'' with the quantum state, insofar as the inner product of
$f$ with the functions $g_i$ that are used to construct the state is a measure of similarity
between the on-shell fourier components of the functions.
There is of course a minimal ``resonance'' of $f$ with vacuum state fluctuations.
In the nonlinear case, in the same folk terms, the nonlinear inner product is a measure of
similarity between not only the on-shell components of $f$ and $g_i$, but also between the
on-shell components of $f^2$ and $g_i^2$, $f+f^2$ and $g_i+g_i^2$, etc.
We cannot, therefore, just add the results of measuring $\hat\phi_f$ and $\hat\phi_g$ to
compute what we would have observed if we had measured $\hat\phi_{f+g}$, because the nonlinear
resonances are not taken into account by simple addition of the operators.

Analogously to equations (\ref{vacuumProb}) and (\ref{vacuumplus1Prob}), we can construct the
pseudoprobabilities
\begin{eqnarray}
  &&\frac{e^{-\Half \underline{x}^T F^{-1}\underline{x}}}
       {\sqrt{(2\pi)^n \mathrm{det}(F)}}, \\
  &&\left[|\underline{x}^T F^{-1}\underline{S}|^2+(1-\underline{S}^\dagger F^{-1}\underline{S})\right]
       \frac{e^{-\Half \underline{x}^T F^{-1}\underline{x}}}
            {\sqrt{(2\pi)^n \mathrm{det}(F)}},\\
  &&F_{ij}=\mathrm{Re}\left[\xi(f_i,f_j)\right],\qquad S_i=\frac{\xi(f_i,g)}{\sqrt{\xi(g,g)}}
\end{eqnarray}
in which the only change, predictably enough, is that we replace the inner product $(f,g)$ by
the ``nonlinear inner product'' $\xi(f,g)$ wherever it occurs.
The probability densities generated for the vacuum state are still Gaussian (which will be addressed
by the method of the next section), but, for example, the fall-off of the 2-measurement correlation
coefficient with increasing distance is controlled by $\xi(f,g)$, so the fall-off is in general
nontrivially different from the fall-off for the free field.
For scalar functions $f(x)$ and $f_a(x)=f(x+a)$ representing two measurements at separation $a^\mu$,
and supposing the dynamics is described by the inner product (\ref{ScalarInnerProduct}) with masses
$m_i$, then the 2-measurement correlation function is given by
\begin{eqnarray}
  \xi(f,f_a)&=&\hbar\sum_i\int\widetilde{\mathcal{P}_i[f]}^*\widetilde{\mathcal{P}_i[f_a]}
                    2\pi\delta(k^\mu k_\mu-m_i^2)\theta(k_0)\frac{\Intd^4k}{(2\pi)^4}\cr
  &=&\hbar\sum_i\int\left|\widetilde{\mathcal{P}_i[f]}\right|^2 e^{-ik_\mu a^\mu}
                    2\pi\delta(k^\mu k_\mu-m_i^2)\theta(k_0)\frac{\Intd^4k}{(2\pi)^4},
\end{eqnarray}
so with a suitable choice of $\mathcal{P}_i$, we have considerable control over the change of the
2-measurement correlation with increasing separation and for different functions $f$.

\section{Weakened linearity II}\label{WeakenedLinearity2}
If we observe non-Gaussian probability densities, we can model them in linear quantum field theory
by acting on the vacuum state with as many creation operators as necessary, spread over as large
a region of space-time as necessary, or by constructing representations of the weakened linear
commutation relations that are unitarily inequivalent to the vacuum sector.
This section discusses the nonlinear alternative already introduced in the introduction,
which maps the creation and annihilation operators nonlinearly to the quantum field $\hat\phi_f$,
\begin{equation}
    \hat\phi_f=\hat F(\hat a^{ }_f+\hat a_f^\dagger,\hat a^{ }_{\mathcal{P}_1[f]}+\hat a_{\mathcal{P}_1[f]}^\dagger,
      \hat a^{ }_{\mathcal{P}_2[f]}+\hat a_{\mathcal{P}_2[f]}^\dagger, X_1(f),X_2(f), ...),
\end{equation}
where $\mathrm{Supp}(\mathcal{P}_i[f])\subseteq\mathrm{Supp}(f)$, and $X_i(f)$ are arbitrary
Poincar\'e invariant scalar functions of $f$.
Microcausality is preserved, $[\hat\phi_f,\hat\phi_g]=0$ whenever $f$ and $g$ have
space-like separated supports, because
$[\hat a^{ }_{\mathcal{P}_i[f]}+\hat a_{\mathcal{P}_i[f]}^\dagger,
  \hat a^{ }_{\mathcal{P}_j[g]}+\hat a_{\mathcal{P}_j[g]}^\dagger]=0\ \forall\hspace{0.1em}i,j$, but if
$\hat F$ includes a dependency on $(f,f)$, for example, there is a larger sense in which the
algebra of observables is nonlocal.
We take the set of observables to be the subalgebra of the algebra of operators generated by
$\hat a^{ }_f$ and $\hat a_f^\dagger$ that is generated by $\hat\phi_f$ (as noted above, the set of observables
in the linear free field case is generated by $\hat\phi_f$, not by the creation and annihilation
operators).

In the simplest case, we can set $G(\hat\phi_f)=\hat a^{ }_f+\hat a_f^\dagger$ for some invertible
function $G(x)$; with this deformation, the gaussian probability density
$Pr(\hat a^{ }_f+\hat a_f^\dagger=x)=\exp{(-x^2/2(f,f))}/\sqrt{2\pi(f,f)}$ becomes
\begin{equation}
  Pr(\hat\phi_f=y)=\frac{1}{\sqrt{2\pi(f,f)}}\exp{\left(-\frac{G(y)^2}{2(f,f)}\right)}G'(y).
\end{equation}
This simplest case is of course more-or-less trivial, but in the most general case the nonlinear
map $F$ is not so easily dismissed.
Whether trivial or not, even for $G(x)=x-\mathrm{tanh}\,x$ we obtain a probability density with
the double maximum characteristic of symmetry breaking,
\begin{equation}
  Pr(\hat\phi_f=y)=\frac{1}{\sqrt{2\pi(f,f)}}\exp{\left(-\frac{(y-\mathrm{tanh}\,y)^2}{2(f,f)}\right)}
   (1-\mathrm{sech}^2\,y)
\end{equation}
(however this is not enough to claim that such a state corresponds to conventional symmetry breaking).
Calculating $n$-measurement correlation functions in this superficially simple model for $n\ge 2$ is
not straightforward.

We have effectively constructed a class of quantum fields that is analogous to the class of integrable
systems in classical field theory in that they are reducible to a free quantum field by nonlinear
(and possibly microcausality preserving but otherwise nonlocal) maps.
In other attempts to construct algebras of observables using the nonlinear operator-valued map
$\hat\phi:f\rightarrow\hat\phi_f$, using algebra deformations similar to those of Arik-Coons
type\cite{Quesne} (which work nicely in the one-dimensional case), I have not so far found it
possible to construct quantum field algebras that are both microcausal and associative, which
I have taken to be essential requirements.

\section{Deformation of electromagnetism}\label{EMfield}
The electromagnetic potential and Dirac spinors are not observable fields, so we will here
deform the quantized electromagnetic field.
To avoid excessive complexity, we will use only the method of section \ref{WeakenedLinearity1}.
The dynamics of the electromagnetic field in terms of a positive semi-definite inner product
on test functions is given by Menikoff and Sharp\cite[equation (3.27)]{MenikoffSharp}
(except for a missing factor of $(2\pi)^{-3}$ that is present in their equation (3.25)):
\begin{eqnarray}
     (f_1,f_2)_{EM} &=& \hbar\int\frac{\Intd^4k}{(2\pi)^4}
             2\pi\delta(k_\alpha k^\alpha)\theta(k_0)
             k^\mu\tilde f_{1\mu\beta}^*(k) k^\nu\tilde f_{2\ \nu}^{\ \beta}(k).
\end{eqnarray}
Note that $f_{1}$ and $f_{2}$ are \emph{not} electromagnetic field tensors, they are classical
test functions that contribute to a description of measurement and/or state preparation of the
quantized electromagnetic field.
The electromagnetic field in an interacting theory of the sort introduced here is not
measurable at a point, so we always have to consider $\hat\phi_f$.

Supposing there is an observable 4-current field, and that $J_{1\mu}$ and $J_{2\mu}$ are
test functions for it, we can introduce a massive free field inner product
\begin{eqnarray}
  (J_1,J_2)_V &=& \hbar\int\frac{\Intd^4k}{(2\pi)^4}
             2\pi\delta(k_\alpha k^\alpha-m^2)\theta(k_0)
\left(\sigma_T k^\mu k_\nu-\sigma_S m^2\delta^\mu_\nu\right)
                     \tilde J_{1\mu}^*(k)\tilde J_2^\nu(k),\cr&&
\end{eqnarray}
where $\sigma_T\ge\sigma_S\ge 0$ determine the relative significance of time-like and
space-like components (relative to $k_\mu$) of the 4-current.
Note that any test function component for which $(f,f)$ is zero is in effect infinitely suppressed
in the free theory\footnote{The variance associated with the observable $\hat\phi_f$ in the vacuum
state is $(f,f)$; if this is zero, then the observed value of $\hat\phi_f$ is always zero in
the vacuum state (and indeed in every state).}, so
$\sigma_S=\sigma_T$ makes only components orthogonal to $k_\mu$ significant and
$\sigma_S=0$ makes only the component parallel to $k_\mu$ significant. 
In terms of these free field inner products, we can introduce an interacting nonlinear inner
product,
\begin{eqnarray}
    &&(\;(J_1,f_1),(J_2,f_2)\;)_I = (f_1,f_2)_{EM}+(J_1, J_2)_V+\cr
    &&\qquad\lambda_1(J_1^\alpha+\kappa_1 J_{1\mu} f_1^{\mu\alpha},
                              J_2^\beta+\kappa_1 J_{2\nu} f_2^{\nu\beta})_V\
                +\lambda_2(J_{1\mu} f_1^{\mu\alpha},J_{2\nu} f_2^{\nu\beta})_V+\cr
    &&\qquad\lambda_3(\epsilon^{\mu\rho\sigma\alpha} J_{1\mu} f_{1\rho\sigma},
                              \epsilon^{\nu\tau\upsilon\beta} J_{2\nu} f_{2\tau\upsilon})_V
\end{eqnarray}
with $\lambda_1$, $\lambda_2$, and $\lambda_3$ all $\ge 0$, and of course higher order terms
are possible.
Degrees of freedom that make no contribution to a noninteracting inner product may
make a contribution after we introduce a new term to a nonlinear inner product.
Fourier components of $J_1$ that are not on mass-shell, for example, so that they make no
contribution to $(J_1,J_2)_V$, may contribute to the on mass-shell fourier components
of $J_{1\mu} f_1^{\mu\alpha}$.
Introducing nonlinearity in this way, therefore, effectively adds new degrees of freedom as well.

Polynomial invariants in derivatives of both $J$ and $f$ can also be added, such as
$(J_1^\alpha+\kappa_2 \partial_\mu f_1^{\mu\alpha}, J_2^\beta+\kappa_2 \partial_\nu f_2^{\nu\beta})_V$
or $(\partial_{[\alpha} J_{1\mu]}+\kappa_3 f_{1\alpha\mu},
  \partial_{[\beta} J_{2\nu]}+\kappa_3 f_{2\beta\nu})_{EM}$,
again with higher orders as necessary.

All the nonlinear terms introduced above can result in correlations between the current and
the electromagnetic field.
In the noninteracting case, the inner product between test functions $(J_1,0)$ and $(0,f_2)$ will
always be zero, so there is no correlation in the vacuum state between 4-current observables and
electromagnetic field observables, but with the introduction of the nonlinear terms above
there will generally be correlations between 4-current observables and electromagnetic field
observables in the vacuum state.
Such interactions between the 4-current and the electromagnetic field
through the action of nonlinearity in this approach are not immediately comparable to the
description of correlations in conventional perturbation theory through the annihilation and
creation of photon and charge lines in Feynman diagrams.

If there is also an observable axial 4-vector, and $S_{1\mu}$ and $S_{2\mu}$ are
test functions for it, quite a few more terms become possible in a nonlinear inner product,
even without introducing derivatives,
\begin{eqnarray}\label{bigI}
    &&(\;(J_1,S_1,f_1),(J_2,S_2,f_2)\;)_I = (f_1,f_2)_{EM}+(J_1, J_2)_V+(S_1,S_2)_V+\cr
    &&\qquad\lambda_1(J_1^\alpha+\kappa_1 J_{1\mu} f_1^{\mu\alpha},
                              J_2^\beta+\kappa_1 J_{2\nu} f_2^{\nu\beta})_V\
                +\lambda_2(J_{1\mu} f_1^{\mu\alpha},J_{2\nu} f_2^{\nu\beta})_V+\cr
    &&\qquad\lambda_3(\epsilon^{\mu\rho\sigma\alpha} J_{1\mu} f_{1\rho\sigma},
                              \epsilon^{\nu\tau\upsilon\beta} J_{2\nu} f_{2\tau\upsilon})_V
                +\lambda_4(S_{1\mu} f_1^{\mu\alpha},S_{2\nu} f_2^{\nu\beta})_V+\cr
    &&\qquad\lambda_5(S_{1\mu} f_1^{\mu\alpha}+
                      \kappa_2\epsilon^{\mu\rho\sigma\alpha} J_{1\mu} f_{1\rho\sigma},
                  S_{2\nu} f_2^{\nu\beta}+
                      \kappa_2\epsilon^{\nu\tau\upsilon\beta} J_{2\nu} f_{2\tau\upsilon})_V+\cr
    &&\qquad\lambda_6(S_{1[\mu}J_{1\alpha]}+
                       \kappa_3\epsilon_{\mu\alpha}^{\ \ \rho\sigma} f_{1\rho\sigma},
                  S_{2[\nu}J_{2\beta]}+
                       \kappa_3\epsilon_{\nu\beta}^{\ \ \tau\upsilon} f_{2\tau\upsilon})_{EM}+\cr
    &&\qquad\lambda_7(S_{1[\mu}J_{1\alpha]},S_{2[\nu}J_{1\beta]})_{EM}
\end{eqnarray}
To these might also be added parity violating terms, and, with the introduction of a scalar inner
product, terms involving $(J_{1\mu}J_1^\mu,J_{2\nu}J_2^\nu)_S$,
$(S_{1\mu}S_1^\mu,S_{2\nu}S_2^\nu)_S$, $(J_{1\mu}S_1^\mu,J_{2\nu}S_2^\nu)_S$,
$(f_{1\mu\alpha}f_1^{\mu\alpha},f_{2\nu\beta}f_2^{\nu\beta})_S$.
Furthermore, every occurrence of an inner product could be modified to make each term have
a unique mass (and a different contribution for the time-like and space-like components of
each 4-current and axial 4-vector term).

In view of the number of parameters that are apparently possible in this approach, even in the
case of electromagnetism, in contrast to the relatively tight constraints imposed by
renormalizability, equation (\ref{bigI}) presumably has to be regarded as only (potentially)
phenomenologically descriptive, not as a fundamental theory, unless a theoretically natural
constraint on admissible terms emerges.
Note that this approach or some extension or modification of it might be empirically useful,
for example if it can describe electromagnetic fields in nonlinear materials effectively,
without it being at all equivalent to QED.

\section{Conclusion}
With all computations being entirely finite, it may be possible to use these nonlinear
quantum field models more easily and with less conceptual uncertainty than using 
conventional perturbation theory.
The universal use of Schwartz space test functions to describe measurement and state preparation
ensures that there are none of the infinities that usually emerge in perturbative quantum field
theory.
Correlation functions for measurements in a given state are straightforwardly computed in
terms of the nonlinear inner products between all the functions used to generate a state
and to describe measurements.

The mathematics allows a reasonable understanding of the nonlinearity that has been introduced,
and there seems to be no \textit{a priori} reason to exclude this kind of nonlinearity, in which the
linearity of the algebra is preserved.
Indeed, on the classical precedent, nonlinearity ought to be expected.
The apparent introduction of nonlinearity by renormalization through the implicit nonlinear
use of test functions gives a stronger impetus to consider how empirically effective the nonlinear
models introduced here --- and perhaps more general models --- can be.

The infinite range of possibilities is at present a little uncontrolled, and the mathematical analysis
of the empirical consequences of particular terms in models of the theory appear to be quite
nontrivial --- to my knowledge it is a novel mathematical problem.
Quantum theory has largely moved to supersymmetry and string theory because of the apparent
impossibility of putting interacting quantum field theory on a sound mathematical footing,
but the form of interacting quantum field theories presented here is a mathematically
reasonable alternative.

It will be interesting to see what range of physical situations can be modelled with these
nonlinear quantum fields.
Free fields are already useful as a first approximation in quantum optics, so it's
possible that the methods of this paper might make a useful second approximation as a
way to construct phenomenological models for nonlinear materials.
These nonlinear quantum fields, however, are conceptually significantly different
from the interacting quantum fields of conventional perturbation theory, and are manifestly
different from conventional constructive and axiomatic quantum fields.

\end{document}